\documentclass[%
 reprint,
 amsmath,amssymb,
 aps,
 prl
]{revtex4-2}

\usepackage{graphicx}% Include figure files
\usepackage{dcolumn}% Align table columns on decimal point
\usepackage{bm}% bold math
\usepackage[dvipsnames]{xcolor}
\usepackage{siunitx}
\usepackage{hyperref}
\usepackage{braket}

\usepackage[dvipsnames]{xcolor}

\usepackage{hyperref}
\newcommand{\um}{$\mu$m}
\newcommand{\upstate}{$|{\uparrow}\rangle$}
\newcommand{\downstate}{$|{\downarrow}\rangle$}
\newcommand{\twostate}{$|{\tilde{\uparrow}}\rangle$}
\newcommand{\Er}{$E_\mathrm{R}$}
\newcommand{\kB}{$k_\mathrm{B}$}

\begin{document}

\preprint{APS/123-QED}

\title{A two-dimensional programmable tweezer array of fermions}% Force line breaks with \\
\author{Zoe Z. Yan$^{1,*}$, Benjamin M. Spar$^{1,*}$, Max L. Prichard$^1$, Sungjae Chi$^1$, Hao-Tian Wei$^{2,3}$, Eduardo Ibarra-Garc\'{i}a-Padilla$^{2,3}$, Kaden R. A. Hazzard$^{2,3}$, Waseem S. Bakr$^1$}

\affiliation{$^1$ Department of Physics, Princeton University, Princeton, New Jersey 08544, USA\\
$^2$ Department of Physics and Astronomy, Rice University, Houston, Texas 77005, USA\\
$^3$ Rice Center for Quantum Materials, Rice University, Houston, Texas 77005, USA}

\date{\today}% It is always \today, today,

\begin{abstract}
We prepare high-filling two-component arrays of up to fifty fermionic atoms in optical tweezers, with the atoms in the ground motional state of each tweezer. Using a stroboscopic technique, we configure the arrays in various two-dimensional geometries with negligible Floquet heating. Full spin- and density-resolved readout of individual sites allows us to post-select near-zero entropy initial states for fermionic quantum simulation. We prepare a correlated state in a two-by-two tunnel-coupled Hubbard plaquette, demonstrating all the building blocks for realizing a programmable fermionic quantum simulator.
\end{abstract}

\maketitle

\begin{figure*}[t]
\includegraphics[width=\textwidth]{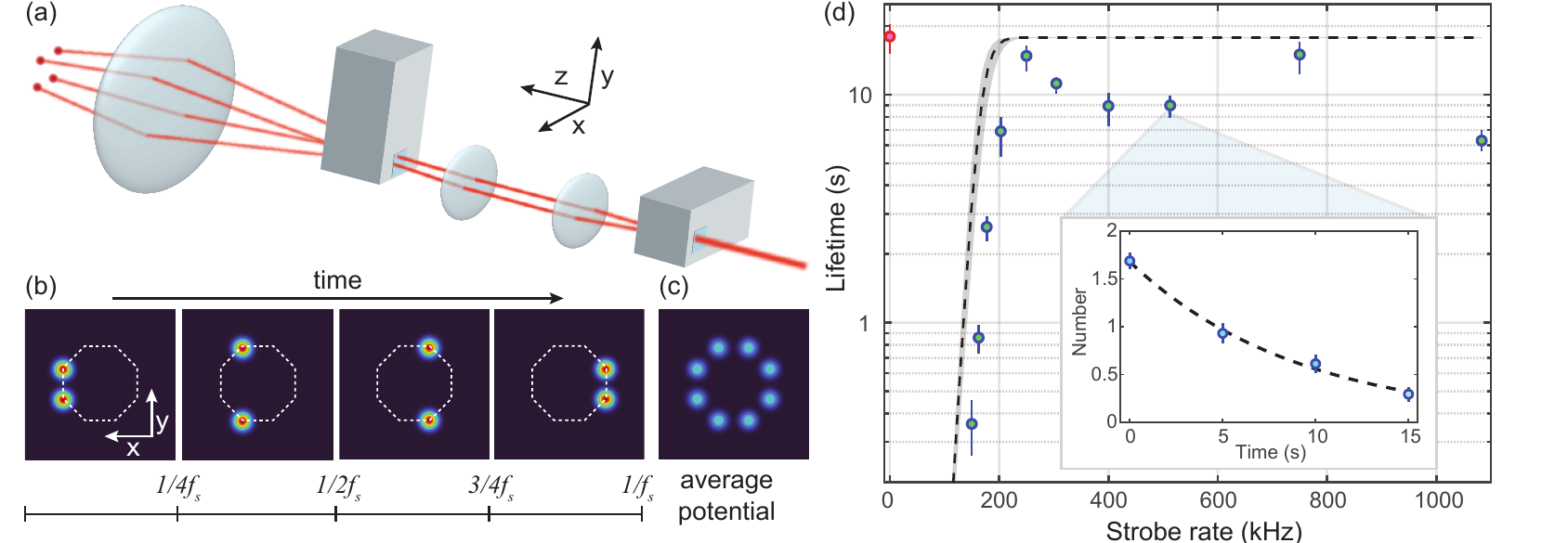}
\caption{\label{fig:1} 2D stroboscopic tweezer technique and lifetimes. (a) Two crossed acousto-optic modulators spaced in a 4$f$ configuration generate the array.  (b) Illustration of the principle of stroboscopic array generation of an 8-site ring.  For a strobe frequency $f_s$, each column of the array is turned on for a quarter of the period $1/f_s$, generating a time-averaged potential shown in (c).  (d) Lifetime of an atom in the ground vibrational state of a tweezer versus strobe frequency, with the red point at 0\,kHz indicating the non-strobed lifetime. The dashed line shows the theoretical prediction, and grey shading indicates the systematic uncertainties on the tweezer waist.  The inset shows an example of a decay curve of population in the ground state for $f_s\,{=}\,513$\,kHz with an exponential fit.}
\end{figure*}

Ultracold atoms in optical tweezer arrays have become a popular platform for quantum simulation, computation, and metrology~\cite{NatPhys_Kaufman_Ni}.
The tweezer platform has recently witnessed rapid breakthroughs, ranging from the development of precise optical clocks~\cite{madjarov2019,young2020half} to the demonstration of entangling operations~\cite{graham2019,levine2019parallel, omran2019generation,madjarov2020high}.
The realization of defect-free arbitrary geometries~\cite{endres2016atom,barredo2016atom}, in particular in two dimensions, has paved the way for studying rich quantum many-body physics with localized Rydberg atoms, including frustrated spin models on a triangular lattice~\cite{Lienhard2018, scholl2021quantum}, topological phases in a zig-zag chain~\cite{deleseleuc2019}, and quantum spin liquids with atoms placed on the links of a kagome lattice~\cite{semeghini2021probing}.

The versatility of tweezer arrays has also been extended to systems of itinerant atoms where quantum statistics play a role~\cite{kaufman2014two,murmann2015two, bergschneider2019, Becher2020, Spar2021,Young2022}. In particular, tunnel-coupled arrays have been realized for small systems of bosonic~\cite{kaufman2014two} and fermionic~\cite{murmann2015two, bergschneider2019, Becher2020, Spar2021} atoms in one dimensional arrays. If these experiments can be scaled, they would constitute a bottom-up approach toward quantum simulation that complements optical lattice experiments with quantum gas microscopes, which currently lie at the forefront of studying one- and two-dimensional Hubbard models~\cite{cheuk2016observation,parsons2016site,boll2016spin, brown2017spin,nichols2019spin, brown2019bad,vijayan2020time,ji2021dynamical,koepsell2019Imaging,bohrdt2021exploration}. The difficulty of reconfiguring microscope experiments has led to an almost exclusive focus on physics in square lattices (Ref.~\cite{yang2021site} is a recent exception). Programmable Hubbard tweezer arrays would allow the extension of site-resolved studies to arbitrary lattice geometries that bring additional ingredients into play, including frustration, topology, and flat-band physics.

Hubbard tweezer arrays also have the potential to address another major challenge for optical lattice experiments: the preparation of low-entropy phases of fermions. In optical lattice experiments, the entropy of the gas is limited by evaporative cooling, which is hindered by poor efficiencies at low temperatures. Entropy redistribution schemes relying on the flow of entropy away from gapped phases have been proposed~\cite{Bernier2009,Lubasch} and experimentally explored~\cite{chiu2018quantum}, but they have not resulted in significant reduction of achieved temperatures. 

Here we show that stroboscopic optical tweezer arrays can be used to prepare low entropy fermionic systems with arbitrary two-dimensional (2D) geometry. The low entropy is possible due to several features particular to this platform. First, in loading a tweezer from a degenerate Fermi gas, the tweezer acts as a ``dimple trap," wherein the local Fermi temperature ($T_\mathrm{F}$) is significantly higher than in the bulk gas. 
Since the fraction of atoms loaded into the tweezers is low, the temperature of the system remains approximately fixed to the bulk gas's temperature, but the tweezers' phase space density is enhanced. Furthermore, the occupancy of the lowest level of each tweezer (given by the Fermi-Dirac distribution) is close to unity. 
This enables the preparation of a state with two atoms in the ground motional state (one per spin state) on every tweezer with high fidelity, as first demonstrated in Ref.~\cite{ serwane2011deterministic}. Second, the system can be evolved from the band insulator into a correlated state via an adiabatic ramp-on of additional sites, taking advantage of independent tunability of each lattice site. We have previously shown that this technique can be used to prepare a state with antiferromagnetic correlations in an eight-site Fermi Hubbard chain~\cite{Spar2021}. We extend this approach to 2D and show that any pre-ramp entropy in the system can be effectively eliminated by post-selection on the atom number in each spin state. Post-selection is enabled by spin- and density-resolved readout~\cite{boll2016spin,koepsell2020robust}, which we implement in a bilayer imaging scheme.

\begin{figure}[b]
\includegraphics[width = \columnwidth]{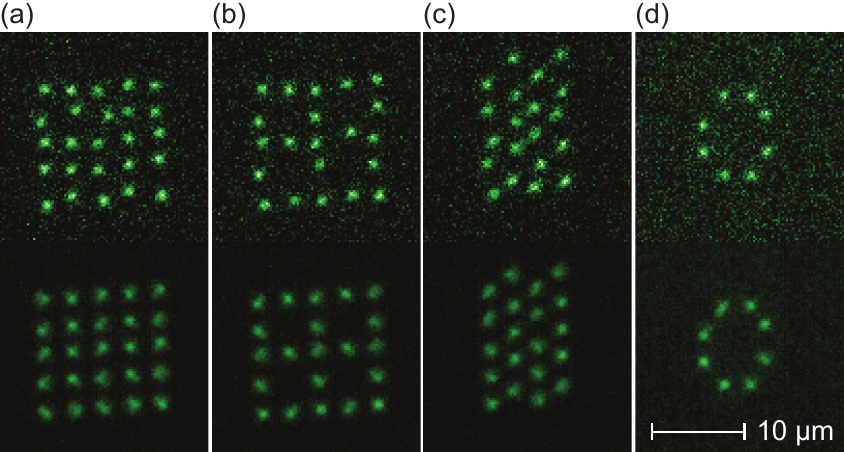}
\caption{\label{fig:2}
Examples of band insulators of different geometries, showing (a) rectangular $5\times5$, (2) 21-site Lieb plaquette, (3) $4\times5$ triangular, and (4) octagonal ring arrays. Only \upstate atoms are imaged and the sites here are not tunnel-coupled. The top row shows single shots with perfect filling of the \upstate~state, and the bottom row shows average images.  Deviations of the atom positions in the single-shot images are due to quantization onto the lattice for imaging. Average fillings of \upstate~ are (93, 92, 91, 89)\%, accounting for imaging fidelity of 98.5\%, out of (411, 254 , 275, 100) shots.}
\end{figure}

The experimental cycle, including tweezer loading, is the same as detailed in Ref.~\cite{Spar2021}.
Tweezers are loaded from a bulk Fermi gas at $T/T_{\rm F}\,{\approx}\,0.2$ that is a balanced mixture of the lowest and third lowest hyperfine ground states (\upstate,~\downstate, respectively). 
Our scheme for generating 2D arrays uses two crossed acousto-optical modulators [Fig.~\ref{fig:1}(a)].
The tweezers are produced using light with a wavelength of $780$\,nm, and their waist at the atoms is $1000^{+180}_{-140}$\,nm.
Radiofrequency tones for both AOMs are generated by a two-channel arbitrary waveform generator, with a tone separation of 8\,MHz corresponding to a tweezer spacing of 1350\,nm in the atom plane.
The aperture size and bandwidth of the modulators currently limit us to ${\sim}\,9$ tweezers in each direction.
The beat frequency of neighboring tweezers is $>100$ times larger than typical tweezer depths, leading to negligible parametric heating.

Homogenizing the tweezer depths is particularly challenging for 2D arrays generated using crossed AOMs.  
A common approach used in Rydberg tweezer experiments is to apply a static set of frequencies consisting of $n_x$ and $n_y$ tones for the $x$- and $y$- directional AOMs, respectively.
This generates a rectangular array of $n_xn_y$ sites; however, the $n_x\,{+}\,n_y$ degrees of freedom from the signal strength of each tone are insufficient to independently tune the depth of each tweezer.
Better homogeneity can be achieved by tuning the relative phases of the tones, but the typical resultant inhomogeneity still exceeds 1\%.
Tunnel-coupled arrays have more stringent requirements for homogeneity, since the energy offsets in tweezers of typical depth $\sim h{\times}50\,$kHz must be controlled to within tunneling energies of $\sim h{\times}250\,$Hz, or better than 0.5\%.

To homogenize arrays to within this precision and produce arrays with arbitrary geometry, we introduce a stroboscopic tweezer technique.
We generate the array one column at a time, with different $y-$directional tones applied in every timestep [Fig.~\ref{fig:1}(b)].  
Effectively, the atoms experience a time-averaged potential of concatenated 1D arrays, as long as the strobe rate $f_s$ far exceeds the tweezers' harmonic trap frequencies.
As the typical axial (radial) trap frequencies are around 2.5\,(15)\,kHz, we need strobe rates over an order of magnitude higher to avoid significant Floquet heating of the atoms.

We verify that the stroboscopic scheme is compatible with long lifetimes in the tweezer ground vibrational state with the following study.
We measure the dependence of the lifetime in the lowest vibrational state on $f_s$ in a two-site strobed array, varying the strobe rate from 163-1083\,kHz~\cite{Supplement}.
Higher frequencies are inaccessible due to limitations on the AOM response rate, set by the speed of sound and beam size in the crystal.
We also compare the lifetimes to that of a static (non-strobed) tweezer, which is limited by background gas collisions and off-resonant photon scattering due to the trapping light.
Consistent with expectations, the lowest strobe rates give the shortest lifetimes in the ground state [Fig.~\ref{fig:1}(d)]. Measurements and numerics using a discrete variable representation (DVR) method~\cite{light2000discrete,wall:effective_2015,Supplement} both indicate that Floquet heating decreases exponentially with increasing $f_s$ and is negligible for $f_s\gtrsim 250$\,kHz, although the numerics underestimate the threshold frequency range below which severe heating occurs by ${\sim}\,18\%$.

We demonstrate loading the arrays with band insulators of fermions with high fidelity using the stroboscopic method. These band insulators serve as low entropy initial states for fermionic quantum simulation. As proofs-of-principle, we implement a rectangular $5\times5$ array, 21-site Lieb plaquette, triangular $4\times5$ array, and an 8-site octagonal ring (Fig.~\ref{fig:2})~\cite{Supplement}. The tweezers are homogenized using a density balancing algorithm where the number of required experimental shots is almost independent of the array size~\cite{Spar2021}.
In these examples, the sites are not tunnel-coupled due to the large separations.
Readout is accomplished by transferring the atoms into a 2D lattice of 752\,nm spacing, which oversamples the tweezer array, and performing Raman sideband cooling on the \upstate~atoms after removal of the \downstate~atoms~\cite{brown2017spin, Spar2021} with a detection fidelity of 98.5\%. Throughout these different geometries, the loading fidelity of a single spin averages to 92\%/site, corrected for imaging infidelity, indicating a low entropy of loading in the array. As in previous work~\cite{serwane2011deterministic,Spar2021}, the tweezer depths are chosen so the predominant type of defect in each tweezer is a missing particle rather than an extra one in a higher motional state. In these data, we only measure one of the spin states in a given experimental shot, due to the problem of light-assisted collisions, which necessitates the removal of the other spin state before imaging~\cite{gross2021quantum}.

\begin{figure}[b]
\includegraphics{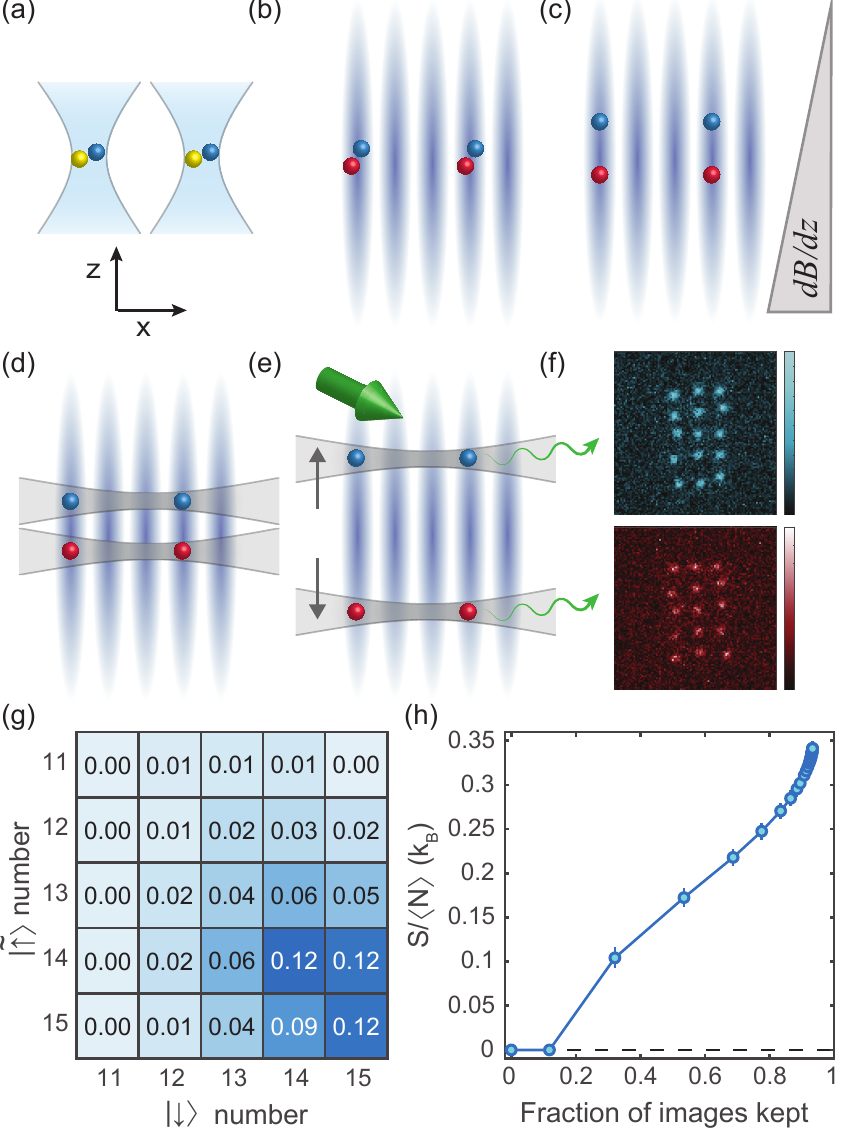}% 
\caption{\label{fig:3} Bilayer imaging procedure and entropy reduction through post-selection.  (a) Atoms in \upstate~(yellow) and \downstate~(blue) are initially trapped in the tweezers, then adiabatically loaded into (b) a 2D lattice with vertical waist of 75\,\um, where \upstate~is transferred to \twostate~(red). (c) A magnetic field gradient is applied to separate the spins in the vertical direction, after which (d) two lightsheet potentials turn on to fix the $z$-positions.  (e) The lightsheets are further separated to 25\,\um~separation.  Raman sideband imaging commences, producing simultaneous images of both spin states.  (f) shows a single shot image of \downstate~and \twostate~originally from a $3\times5$ rectangular array.
(g) Probability distribution versus number of atoms in each spin state over 972 shots. Here, all images with doublons (65 shots) were not used.  (h) By post-selecting on the maximum number of holes, effective entropy can be reduced by varying amounts.  }
\end{figure}

To circumvent this problem and obtain full density- and spin-resolution, we adopt a high-fidelity bilayer imaging scheme~\cite{preiss2015quantum,koepsell2020robust,hartke2020doublon,gall2021competing}, which also allows the reduction of entropy upon post-selection.
Bilayer density- and spin-readout was first accomplished in fermionic quantum gas microscope experiments in a superlattice charge-pumping scheme~\cite{koepsell2020robust}.
Our method is conceptually similar but involves no superlattice (Fig.~\ref{fig:3}).
Starting with tweezer-trapped atoms, we adiabatically turn off the tweezer and turn on a 2D lattice of 1064\,nm to 60\,\Er~and a vertical trap frequency of 1.2\,kHz in 5\,ms.  
The magnetic field is brought to 572\,G, where we perform a spin-flip of \upstate~to the second-lowest hyperfine state, \twostate, with an efficiency exceeding 99\%, and then decrease the field to near 0\,G.  
Atoms in \twostate~and \downstate~have a greater differential magnetic moment than those in \upstate, enabling the Stern-Gerlach separation of these populations to ${\sim}\,9$\um~using a $z$-magnetic gradient of 168 G/cm in the 2D lattice at a depth of 280\,\Er.
We turn on two lightsheet potentials~\cite{brown2017spin}--highly anisotropic beams, each with $z$-directional trap frequencies of 26\,kHz--and linearly ramp their vertical separation to 25\,\um~for imaging.
We measure a combined transport and spin identification fidelity of 98.7\%.
Finally, we image the atoms using Raman sideband cooling simultaneously in both layers, with the 2D lattice depth at 2500\,\Er~and the two lightsheet $z$-trap frequencies at 70\,kHz.
Resulting fluorescence is collected by a high numerical aperture objective with atoms in the two planes focused onto two different active areas of a CCD camera. Imaging fidelity is 98\% (97\%) for the layer of \twostate ~(\downstate) atoms.

Bilayer imaging enables reduction of the effective entropy associated with the initial state of the tweezer array (the band insulator) through post-selection. 
The initial entropy per particle of the tweezer ensemble, assuming independent tweezers and single-band occupation, is given by
\begin{equation}
 \frac{S}{\langle N\rangle} = -\frac{k_\mathrm{B}}{p} \big(p\log p+(1-p) \log (1-p) \big).
\end{equation}
where $p$ is the probability to load one spin on a site.
With a typical loading efficiency of $p\,{=}\,0.907(3)$, the array starts with 0.34(1)\,\kB~per particle, with entropy entering from microstates with undesired holes.
By selecting only images with the population per spin state equal to the number of loading tweezers, we can effectively choose a subsample with $S\,{=}\,0$. Importantly, this post-selection criterion eliminates the initial state entropy even after changing the filling of the system (by introducing additional tweezers) to prepare a correlated state. The post-selection criterion can be relaxed to use more images from the experiment at the cost of introducing additional initial state entropy. This tradeoff is illustrated in Figs.~\ref{fig:3}(g-h) for a $3\times5$ array in which \upstate~and \downstate~had average $p\,{=}\,$0.914(3) and 0.900(3), respectively (not accounting for imaging fidelity).
Out of 972 images, 12\% had perfect filling of 15 fermions of each species. However, even keeping images with up to two holes, or over 50\% of shots, still results in a low entropy of 0.17(1)\,\kB~per particle, which is favorable compared to state-of-the-art optical lattice experiments that range from 0.3{-}0.5\,\kB~per particle~\cite{mazurenko2017cold,brown2019bad,sompet2021realising}.

While post-selection can be used to reduce the effective entropy of the initial state to near zero, subsequent ramps to correlated states will inevitably introduce additional entropy. Numerical simulations of the dynamical ramps in small systems indicate this extra entropy should be low for defect-free initial configurations. For example, for the ramp used in our previous work with an eight-site chain~\cite{Spar2021}, the ramp is expected to have introduced an additional entropy of 0.04\,\kB~per particle when starting with a defect-free state, but the presence of even a single localized hole would lead to a significant entropy increase of 0.2-0.3\,\kB~per particle depending on the position of the hole. The entropy reported in Fig.~\ref{fig:3}(h) should therefore be treated only as a lower bound for future experiments. 

Post-selection on spin and density in this context should be distinguished from the context of optical lattice-based quantum gas microscopy measurements.
For example, in a recent study with a fermionic microscopes~\cite{sompet2021realising}, spin- and density- readout enabled post-selection of half-filled systems with zero total magnetization, keeping ${\sim}\,$9\% of data.  However, post-selection there did not eliminate the finite spin temperature in the initial state.

\begin{figure}[b]
\includegraphics{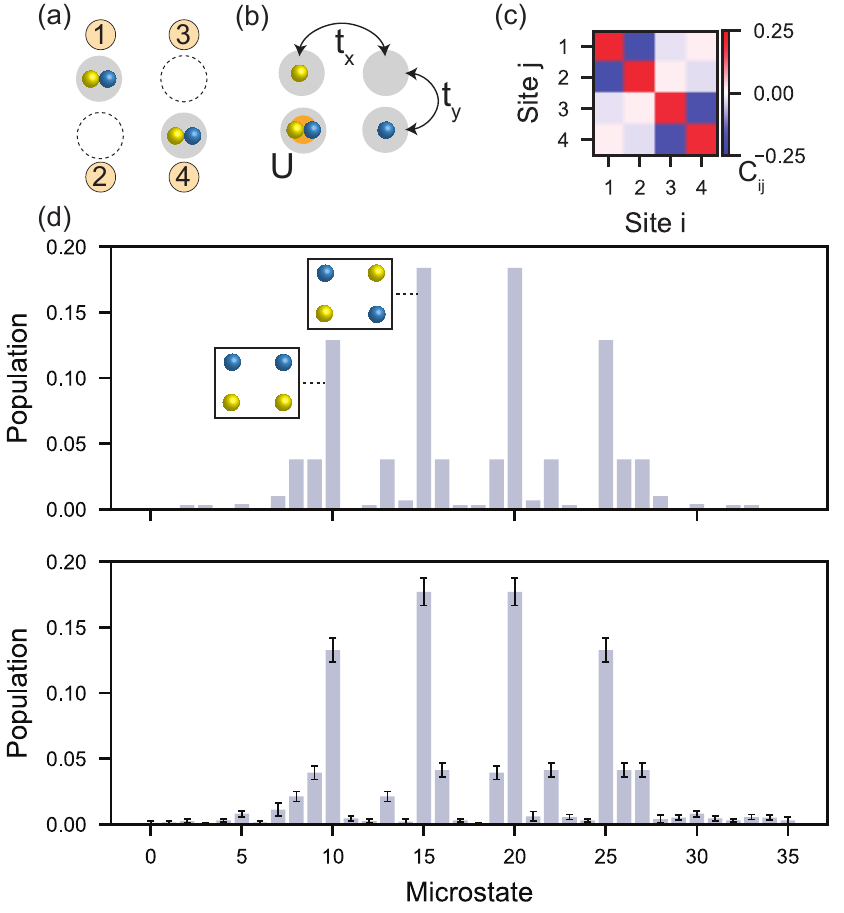}
\caption{\label{fig:4} Low entropy preparation of a $2\times2$ array. a) We load two atoms per site on one diagonal of the array. b) We create a correlated state by ramping on the additional lattice sites and increasing the scattering length to introduce on-site interactions. For the following data, we work with $t_x = 140(5)$ Hz, $t_y = 220(5)$ Hz and $U/\bar{t} = 3.4(2)$. c) Measured spin-spin correlations enabled by the bilayer imaging scheme. d) Best fit (top) and measured (bottom) microstate populations for 671 post-selected experimental shots. The fit gives an entropy in the range [0,0.09] $k_B$ per particle. Insets are shown for the two most common states.}
\end{figure}

Equipped with the ability to load near-zero-entropy band insulators after post-selection, we implement the simplest building block of a two-dimensional Fermi-Hubbard model: a tunnel-coupled $2\times2$ plaquette.
The single-band Hamiltonian is
\begin{align}
    \hat{H} &= - \sum_{\langle i, j\rangle_x, \sigma} t_{x} (\hat{c}^{\dag}_{i \sigma} \hat{c}_{j \sigma} {+} {\rm h.c.})\,{-} \sum_{\langle i, j\rangle_y, \sigma} t_{y} (\hat{c}^{\dag}_{i \sigma} \hat{c}_{j \sigma} {+}{\rm h.c.}) \nonumber\\
    &  \,{+}\, \sum_{i}U_i \hat{n}_{i \uparrow} \hat{n}_{i \downarrow} {+} \sum_{i, \sigma} \Delta_i \hat{n}_{i \sigma},
\end{align}
where $\hat{c}^{\dag}_{i \sigma}$ is the fermionic creation operator of spin $\sigma$ at site $i$, $\hat{n}_{i\sigma}$ is the number operator, $t_{x(y)}$ is the tunneling matrix element in the $x$($y$) direction, $\Delta$ is the energy offset, and $U$ is on-site interaction between opposite spin states.
We start by loading two diagonal sites in a rectangular array with vertical (horizontal) spacing of 1520 (1690) nm [Fig. \ref{fig:4} (a,b)].
The correlated state at half-filling is prepared by adiabatically ramping on the two opposing diagonal sites in 50 ms \cite{Spar2021}, with tunnelings of $t_x\,[t_y] = h{\times}140(5) [220(5)]$\,Hz in the final configuration~\cite{Supplement}. We also ramp $U/\bar{t}$ from 0 to 3.4(2) in the same time using the Feshbach resonance. Here, $\bar{t}\,{=}\,(t_x+t_y)/2$.

The resulting spin-spin correlations are shown in Fig.~\ref{fig:4}(c), which depicts $C_{ij}\,{=} \langle S_{z,i}S_{z,j}\rangle {-}\langle S_{z,i}\rangle\langle S_{z,j}\rangle$, where $S_{z,i}{\equiv}\frac{1}{2}(n_{\uparrow,i}{-}n_{\downarrow,i})$. Here, data were post-selected to include only images that contained two \upstate~and two \downstate~atoms, for a total of 673 experimental cycles. With full spin- and density-readout, we are able reconstruct the diagonals of the density matrix $\rho\,{=}\,|\Psi\rangle\langle \Psi|$ in the basis of allowed number states (with Hilbert space size ${{4}\choose {2}}^2{=}\,36$), and compare data with theory. In Fig.~\ref{fig:4}(d), we plot the experimental population in each microstate together with the populations expected theoretically for the plaquette ground state, which are consistent within error bars. Here, we reduce our statistical errors by taking advantage of the spin-symmetry of the Hubbard Hamiltonian to average the probabilities for spin-reversed microstates. Furthermore, we fit the temperature of the canonical ensemble to best reproduce the distribution of microstates. The fit gives an upper bound for the temperature of $k_B T\sim0.3\bar{t}$ (with the fit losing sensitivity below that temperature). This corresponds to an entropy in the range [0,0.09] $k_B$ per particle, which is consistent with the prediction from simulating the ramp dynamics (entropy gain of 0.02 $k_B$ per particle).  

In conclusion, we have realized a 2D tweezer array of fermions with software-programmable geometry using a novel stroboscopic technique that allows independent control over all tweezer depths and positions. We have realized the building blocks to implement programmable 2D Fermi-Hubbard models, and demonstrated these on a small scale. Future work will focus on increasing the system size of the tunnel-coupled arrays. A natural target for future work will be few-leg ladder systems. For example, two-leg triangular ladder systems can be used to explore the $J_1-J_2$ model, including the special case of the Majumdar-Ghosh model and its valence-bond solid ground states~\cite{girvin2019modern}. Furthermore, upon introducing spin-imbalance and hole-doping, a triangular two-leg ladder is predicted to host magnon-hole binding at energy scales set by the tunneling, rather than the superexchange~\cite{morera2021attraction}. Multi-leg triangular ladders may potentially host other exotic states such as a chiral spin liquid at half filling and intermediate $U/t$ that evolves to a 120$^\circ$-antiferromagnetic order at strong $U/t$~\cite{Szasz2020}. Ultimately, fully 2D tunnel-coupled arrays with arbitrary geometry will be a rich playground for exploring novel phases of correlated fermions.

We would like to thank Grace Sommers and Elmer Guardado-Sanchez for experimental assistance and Martin Zwierlein's group for the loan of equipment.  The experimental work was supported by the NSF (grant no. 2110475), the David and Lucile Packard Foundation (grant no. 2016-65128), and the ONR (grant no. N00014-21-1-2646). K.R.A.H, H.-T.W., and E.I.-G.-P. acknowledge support from the Welch Foundation through Grant No. C1872, the Office of Naval Research Grant No. N00014-20-1-2695, and the National
Science Foundation through Grant No. PHY1848304.  K.R.A.H. also benefited from discussions at the KITP, which was supported in part by
the National Science Foundation under Grant No. NSF
PHY-1748958.

\noindent$^*$ These authors contributed equally to this work.

\bibliography{bib}% Produces the bibliography via BibTeX.

\setcounter{figure}{0}
\setcounter{equation}{0}
\setcounter{section}{0}

\clearpage
\onecolumngrid
\vspace{\columnsep}

\newcolumntype{Y}{>{\centering\arraybackslash}X}
\newcolumntype{Z}{>{\raggedleft\arraybackslash}X}

\newlength{\figwidth}
\setlength{\figwidth}{0.45\textwidth}

\renewcommand{\thefigure}{S\arabic{figure}}
\renewcommand{\theHfigure}{Supplement.\thefigure}
\renewcommand{\theequation}{S\arabic{equation}}
\renewcommand{\thesection}{\arabic{section}}

\begin{center}
    \large{\textbf{Supplemental Information: \\ A two-dimensional programmable tweezer array of fermions}}
\end{center}

\section{Measurement of Floquet heating}\label{sec:lifetime}

To measure the ground state lifetimes shown in Fig.~1 of the main text, we perform the following experiment.
We load two tweezers, spaced 8.1\,\um~apart to avoid any overlap, each with a spin up and a spin down atom. Atoms in higher vibrational states are removed by lowering the tweezer depth and applying a magnetic field gradient, as described in \cite{Spar2021}. The tweezer depth is increased to 50\,kHz, and a variable hold time is applied. The tweezer light is strobed such that only one tweezer is on at a time, resulting in Floquet heating for low strobe frequencies. Finally, atoms in newly populated higher vibrational states are again removed with an identical spilling process, and the remaining \upstate~atoms are imaged.

\section{Theoretical calculations of Floquet heating}

We determine the Floquet heating from the strobing of the trap by directly  calculating the dynamics of a single atom in a tweezer potential  using  a method based on a discrete variable representation (DVR) of Hilbert space~\cite{light2000discrete}, which was applied to ground state properties of tunnel-coupled tweezers in~\cite{wall:effective_2015,kaufman2014two}. The tweezer potential is 
\begin{equation}
 V(\vec{r}) = - \frac{V_0}{1+\frac{z^2}{z_R^2}} \exp \left[- \frac{2r^2}{w_0^2 (1+\frac{z^2}{z_R^2})} \right] \label{eq:trap-pot}
\end{equation}
where $V_0$ is the trap depth when the tweezer is on, $z_R$ is the Rayleigh range, and $w_0$ is the trap waist, and we use the measured parameters. We use a DVR basis of spatial-parity-adapted sinc functions on a three-dimensional cubic grid of $(N_x,N_y,N_z)$ points spanning $(0,0,0)$ to $(L_x,L_y,L_z)$ to represent our wavefunctions. Only even parity states occur for the dynamics considered.

The state at time $t$ after the strobed time evolution   is 
\begin{align}
\ket{\psi(t)} = \underbrace{\left[e^{-i {\hat T} \delta t}e^{-i ({\hat T}+{\hat V}) \delta t}\right]\cdots \left[e^{-i {\hat T} \delta t}e^{-i ({\hat T}+{\hat V}) \delta t}\right]}_{t/2\delta t \text{ times}} \ket{i} \label{eq:strobe-evolve}
\end{align}
where ${\hat T}=-\hbar^2\nabla^2/2m +V_{\text{abs}}(\vec{r})$ is the kinetic energy operator ($\hbar$ is Planck's constant, and $m$ is the mass of $^6$Li) plus a potential ${\hat V}_{\text{abs}}$ to handle boundary conditions discussed below,   ${\hat V}$ is the potential energy operator associated with Eq.~\eqref{eq:trap-pot}, and $\delta t = 1/(2f_s)$, and assuming $t/2\delta t$ is an integer (the micromotion between such times is discussed later). 
The initial state $\ket{i}$ is assumed to be the ground state of the system in the time-averaged potential, i.e. of $H_{1/2}=T+V/2$.  
Directly solving Eq.~\eqref{eq:strobe-evolve} is challenging due to three factors: (1) the tweezer potential is 3D, (2) it involves a wide range of length scales (the wavefunction localization length, the Gaussian waist, and the Rayleigh range), and (3) the dynamics problem must account for a huge separation of timescales when $f_s$ is large, the short time of the pulses and the long lifetime of the atoms, a ratio of timescales of more than $10^7$. 

We solve this by  rewriting Eq.~\eqref{eq:strobe-evolve}. First we can diagonalize to obtain
$    e^{-i {\hat T} \delta t} = U_0  D_0 U_0^{-1}
$
where  $D_0$ is  diagonal and $U_0$ is  the matrix whose columns are eigenvectors of ${\hat T}$, obtained by first diagonalizing $\hat T$ and then  exponentiating. 
($U_0$ need not be unitary since ${\hat T}$ has non-Hermitian terms for the absorbing potential.)  Similarly
$
e^{-i ({\hat T}+{\hat V}) \delta t} = U_1 D_1 U_1^{-1},
$.
Therefore the terms in brackets -- denote it $M \equiv e^{-i {\hat T} \delta t}e^{-i ({\hat T}+{\hat V}) \delta t}$  -- in  Eq.~\eqref{eq:strobe-evolve} can be rewritten
\begin{equation}
M =U_0 D_0 U_0^{-1}   U_1   D_1 U_1^{-1}.
\end{equation}
One can multiply these matrices, and diagonalize the result to find  $M=U_2 D_2 U_2^{-1}$, so  
\begin{equation}
\ket{\psi(t)} = M^{t/2\delta t} \ket{i}  =  U_2 D_2^{t/2\delta t} U_2^{-1}  \ket{i}. 
\end{equation} 
 %assuming that $t$ is a multiple of $1/f_s$. 
 The micromotion at times between integer multiples of $2\delta t$ can be treated by propagating to the largest multiple of $1/f_s$ as above, and then calculating the time evolution within a single period, which is straightforward and efficient in terms of the diagonalized evolution matrices ($D_a$, $U_a$) already obtained.
The computational cost is dominated by diagonalizing to find $D_0$, $D_1$, and $D_2$, with several additional multiplications.
Although this involves full diagonalization of three large matrices, it avoids the issues associated with separation of timescales.  

\begin{figure}[htb!]
    \centering
    \includegraphics[width=0.45\textwidth]{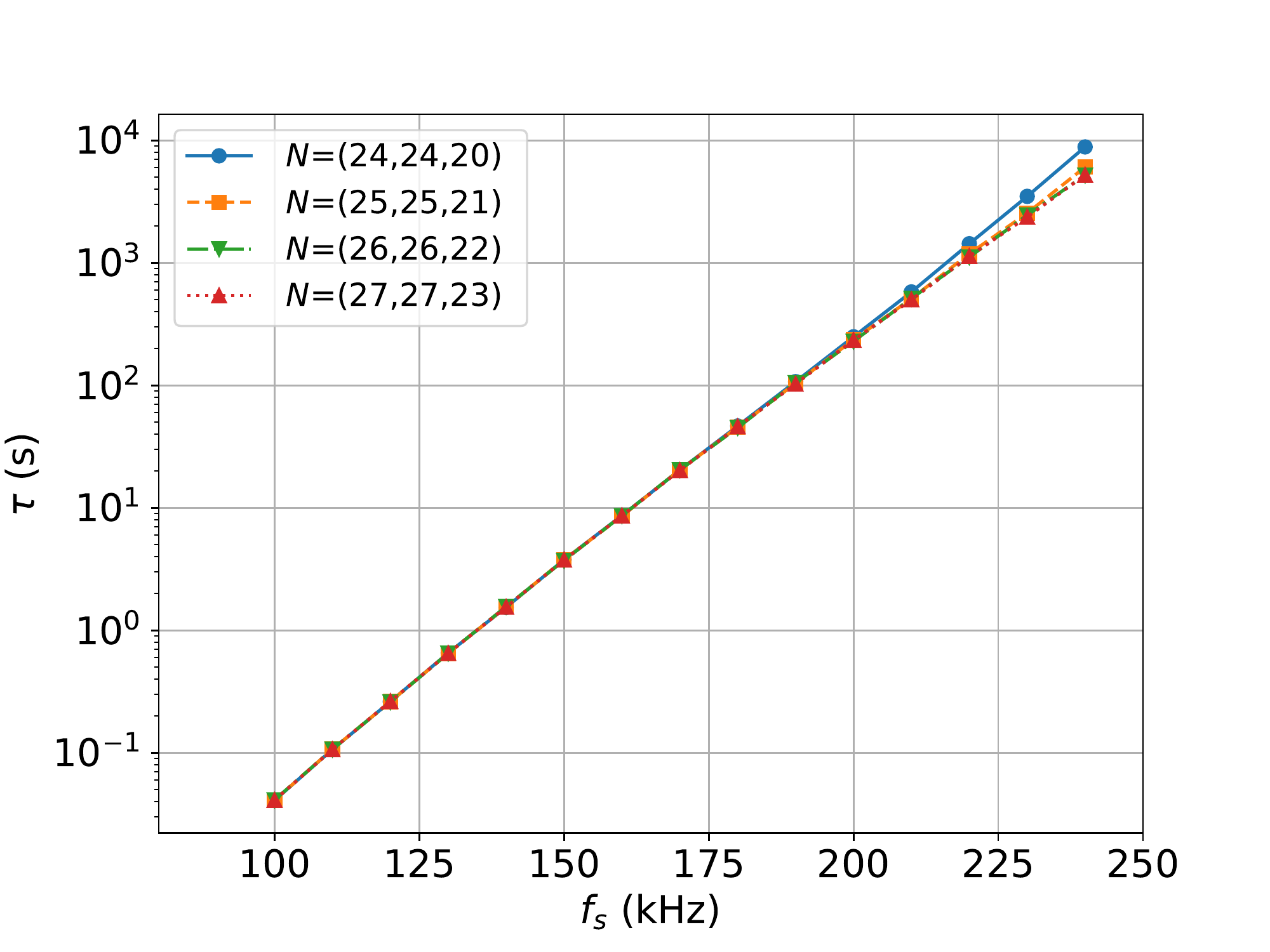}
    \quad
    \includegraphics[width=0.45\textwidth]{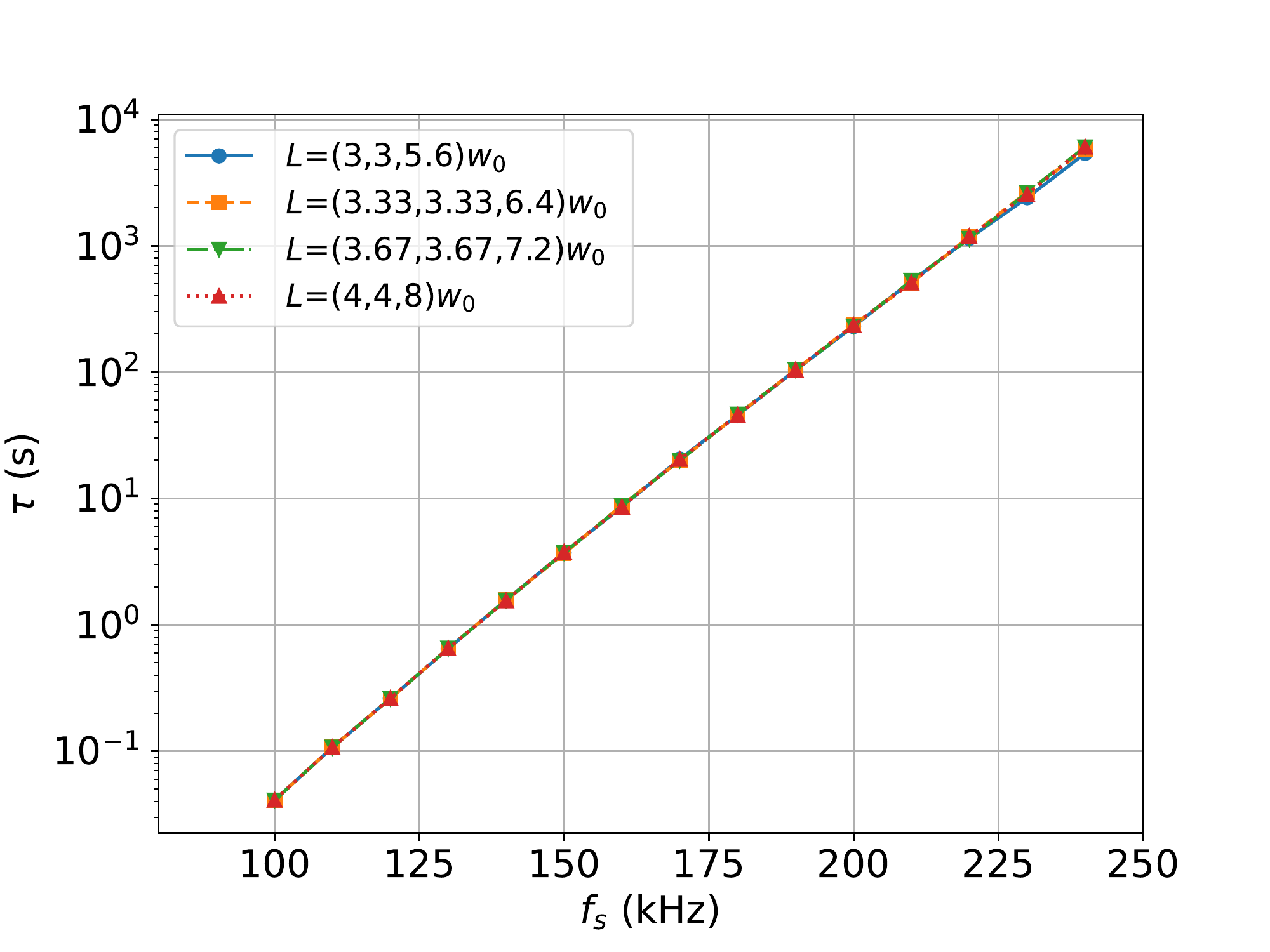}
    \caption{Convergence of the calculated atom lifetime  with respect to number of DVR grid points (left) and system size (right). Left: Lifetimes for different numbers of DVR grid points are converged well in the number of gridpoints for $(N_x,N_y,N_z)=(25,25,21)$ (or more) in  systems of size $(L^{(0)}_x,L_y^{(0)},L_z^{(0)})=(3,3,7.2)w_0 $. Right: lifetimes for different system sizes are converged well for $(L_x,L_y,L_z)=(3,3,5.6)w_0$ and larger. Right panel is at fixed grid point spacing $(0.16667,0.166667, 0.4)w_0$, so the $N_a$ vary with $L_a$. The spacing corresponds to $(N_x,N_y,N_z)=(24,24,20)$ for the largest system.     }
    \label{fig:N&R}
\end{figure}

To obtain accurate results, one must use a a fine enough DVR grid  
%(large enough $N$) 
and a large enough system.
%(large enough $L_x^{(0)}$, $L_y^{(0)}$, $L_z^{(0)}$). 
Here, a challenge  arises  that is absent for ground state calculations: atoms can be excited from the ground state to scattering states and thus escape from the trap. They may move rapidly, and thus would require  $L_a$ that grow linearly with time to capture, in practice many orders of magnitude larger than we can calculate. However, in the experiment once particles move sufficiently far outside the trap, the probability they return is negligible. This can be accurately reproduced  in the calculations by including an absorbing boundary region in our simulations, a standard technique in calculations of, e.g., chemical dynamics~\cite{neuhasuer:time-dependent_1989}. We use a potential 
\begin{equation}
V_{\text{abs}}(\vec{r}) = -i \Gamma \! \sum_{a\in x,y,z}\frac{|a|-L_a^{(0)}}{L_a-L_a^{(0)}}.
\end{equation}
This is zero inside the cubic region from the origin to $(L_x^{(0)},L_y^{(0)},L_z^{(0)})$, and linearly increases as one moves outside this region. The whole calculation is performed inside a box of size $(L_x,L_y,L_z)$ with $(N_x,N_y,N_z)$ grid points. The calculations in the main text are performed with  $(L_x^{(0)},L_y^{(0)},L_z^{(0)})=(3,3,7.2)w_0$,  $N=(27,27,23)$;  $\Gamma=\SI{57}{\kilo\hertz}$; and $L_a=L_a{(0)}+S_a$ where $S_a\approx w_0$, and are  well-converged in the number of grid points and system size (Fig.~\ref{fig:N&R}).  ($S_a$ varies slightly in the right panel: because we fix the  grid spacing and ensure that the last grid point in the non-absorbing region doesn't change with $L_a$, we  choose  $S_a$ closest to $w_0$ while constraining $L_a$ to correspond to the location of the last grid points along the $a$'th direction.)  The $\Gamma$ and $S$ are chosen to be sufficiently large so that particles leaving the trap are absorbed before reaching the boundary of the calculation, but $\Gamma$ is maintained sufficiently small so that there is no Zeno effect  that would reflect the particles. The choices we make are consistent with these conditions, as given in Ref.~\cite{neuhasuer:time-dependent_1989}, and our results are insensitive to the specific choice of the parameters. For example,  varying $\Gamma$ over a range of  a factor of hundred has  a negligible effect on the atom lifetime. Calculated lifetimes  are converged to have visually negligible errors from  the number of grid points, system size, and absorbing boundary.

Using this method, we calculate the population of the ground state  (of the time-averaged potential) as a function of time. We find it gives a nearly perfect exponential decay   for the $f_s$ shown, with small deviations from exponential at smaller $f_s$.
The lifetimes shown in Fig.~1(d) of the main text are obtained by fitting an exponential $ e^{-t/\tau}$ to the obtained dynamics. The numerical calculations do not include the effects that give the finite lifetime in the unstrobed tweezer (background gas collisions and off-resonant light scattering), and which are expected to be the reason for the high-frequency lifetime saturation in Fig.~1(d). To incorporate those effects, the theory curve in Fig.~1(d) is a simple interpolation  $1/\tau_{\text{eff}} = 1/\tau+1/\tau_{\text{static}}$ where $\tau_{\text{static}}$ is the lifetime in the unstrobed tweezer. The plot is largely insensitive to the details of this interpolation.

\section{2D Array Generation}\label{sec:2d}

Our arrays are generated by two perpendicularly oriented AOMs from AA Optoelectronic (MT110-B50A1,5-IR) driven by a two-channel arbitrary waveform generator (Spectrum M4i.6621-x8 PCIe).
The typical strobing scheme, illustrated in Fig.~1 of the main text, generates one column of the array at any given time. 
This gives the most flexibility for creating different geometries.
The two AOMs operate with the longitudinal acoustic mode with a high speed of sound (4200\,m/s) in the crystal.  
This speed, along with the aperture size of $1.5\times2$\,mm$^2$, sets the minimum possible dwell time on a tweezer (the duration for which the tweezer is on). We find that dwell times below 0.5\,$\mu s$ produce tweezers with profiles that are significantly distorted along the strobe axis.
When generating a two-site array by strobing, we measure that the tweezer waist along the strobe axis increases at 1.4\% per increase of 100\,kHz strobing rate. Distortions in the tweezer profile can be partly mitigated by applying a cosine-sum window function on the waveform that is sent to the strobed AOM. Given the minimum dwell time per tweezer, the strobe period $1/f_s$ grows linearly with the number of strobed columns. This seemingly limits us to about 8 sites along the strobe axis due to Floquet heating based on the results shown in Fig.~1 of the main text.
However, we can surpass this limitation with the technique discussed in the following paragraph.

\begin{figure}[htb]
\includegraphics[width = 4 in]{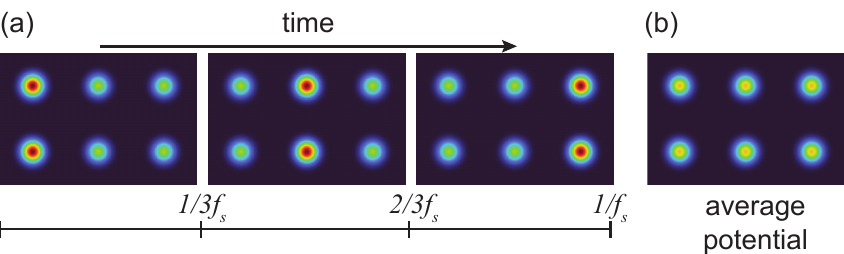}
\caption{\label{fig:s1} Illustration of using 50\% modulation depth to generate a $2\times3$ array with strobe rate $f_s$, with (a) the individual strobe steps and (b) the time-averaged potential.  This procedure can be thought of as superimposing a static $2\times3$ grid on top of a 3-step array with 100\% modulation depth.}
\end{figure}

To reduce Floquet heating and distortion of the tweezer intensity profile, we combine 2D multitone arrays (produced by driving both AOMs with a set of radiofrequency tones) with strobing. For certain geometries, this allows us to reduce tweezer dwell times by using smaller modulation depths. We define modulation depth as the difference between the minimum and maximum relative intensity used over one strobe period. The principle is illustrated in Fig.~\ref{fig:s1} for 50\% modulation depth, and was applied for the rectangular, Lieb, triangular, and ring lattices in Fig.~2 of the main text. For the example of Fig.~\ref{fig:s1}, a static 2D multitone configuration (modulation depth of 0\%) can be used in principle to create the geometry shown, but would not give enough degrees of freedom to homogenize the depths of all sites.

The type of lattices that can be generated with this approach is restricted to combinations of patterns that can be generated by using 2D multitone arrays.
Assume each of the two AOMs is driven by a set of tones such that it individually produces tweezers at positions $P_x, P_y$ respectively, where $P_x\,{=}\,\{p_{x,1},...p_{x,n}\}$ and $P_y\,{=}\,\{p_{y,1},...p_{y,m}\}$.
The result when driving both AOMs will be the Cartesian product of the two sets $P_x\times P_y = \{(p_{x,i},p_{y,j})\,|\,p_{x,i}\in P_x, p_{y,j}\in P_y\}$.
For example, triangular and Lieb lattices can be generated by switching between two multitone rectangular arrays and overlaying a small amplitude strobed array to homogenize the tweezer depths.
In general, most 2D patterns of interest that host some level of periodicity can be generated by switching between two or three multitone rectangular arrays at full modulation depth and overlaying a small amplitude strobed array for homogenization. When using 2D multitone arrays, we avoid square geometries, as equidistant tweezer spacing in $x$ and $y$ leads to diagonal sites having degenerate tones, which results in low frequency beating between these tweezers and atom heating.

\section{$2\times2$ plaquette}

We calibrate the tunnelings in the horizontal and vertical directions in the $2\times2$ array as follows. We first load atoms on only one site, and reduce the tweezer intensities by 5 percent on all other sites except for the one in the direction we want to measure the tunneling. At a non-interacting magnetic field, we quickly initialize tunneling by zeroing the offsets among the target sites~\cite{Spar2021}, and we measure Rabi oscillations between the two sites \cite{murmann2015two,kaufman2014two}.

To calibrate the on-site interaction energy $U$, we perform radiofrequency spectroscopy and measure the difference between single and double occupied sites. However, the most sensitive way to measure $U/t$ is to fit the distribution of microstates. In particular, for temperatures far below the interaction energy, $U/t$ sensitively determines the number of single occupied sites. The value of $U/t$ reported in the main text comes from a best fit of the experimental data, which is in reasonable agreement with an independent spectroscopic calibration [$U/t=4.4(5)$].

To prepare the $2\times2$ plaquette ground state, we initialize the ground state of sites 1 and 4 with doublons, as pictured in Fig.~4 of the main text.  To achieve a half-filled system, we turn on the opposing diagonal sites in a two-step sequence.  
First, the intensities of sites 2 and 3 are increased in 5\,ms with an exponential time constant of 1\,ms, until they are at $\sim$90\% intensity of sites 1 and 4.  
Then, the intensities are further increased in a 50\,ms exponential ramp with a time constant of 5\,ms, until the offsets are zeroed.
During this time, the magnetic field is ramped from a nearly non-interacting system at 573\,G to its final value of 631\,G using a two-point cubic spline function. 
A hold time of 4\,ms is applied for further equilibration.
Finally, tunneling is frozen by rapidly increasing the tweezer intensity to 2.5 times the science depth, and imaging commences.

\begin{figure}[]
\includegraphics[width = 5 in]{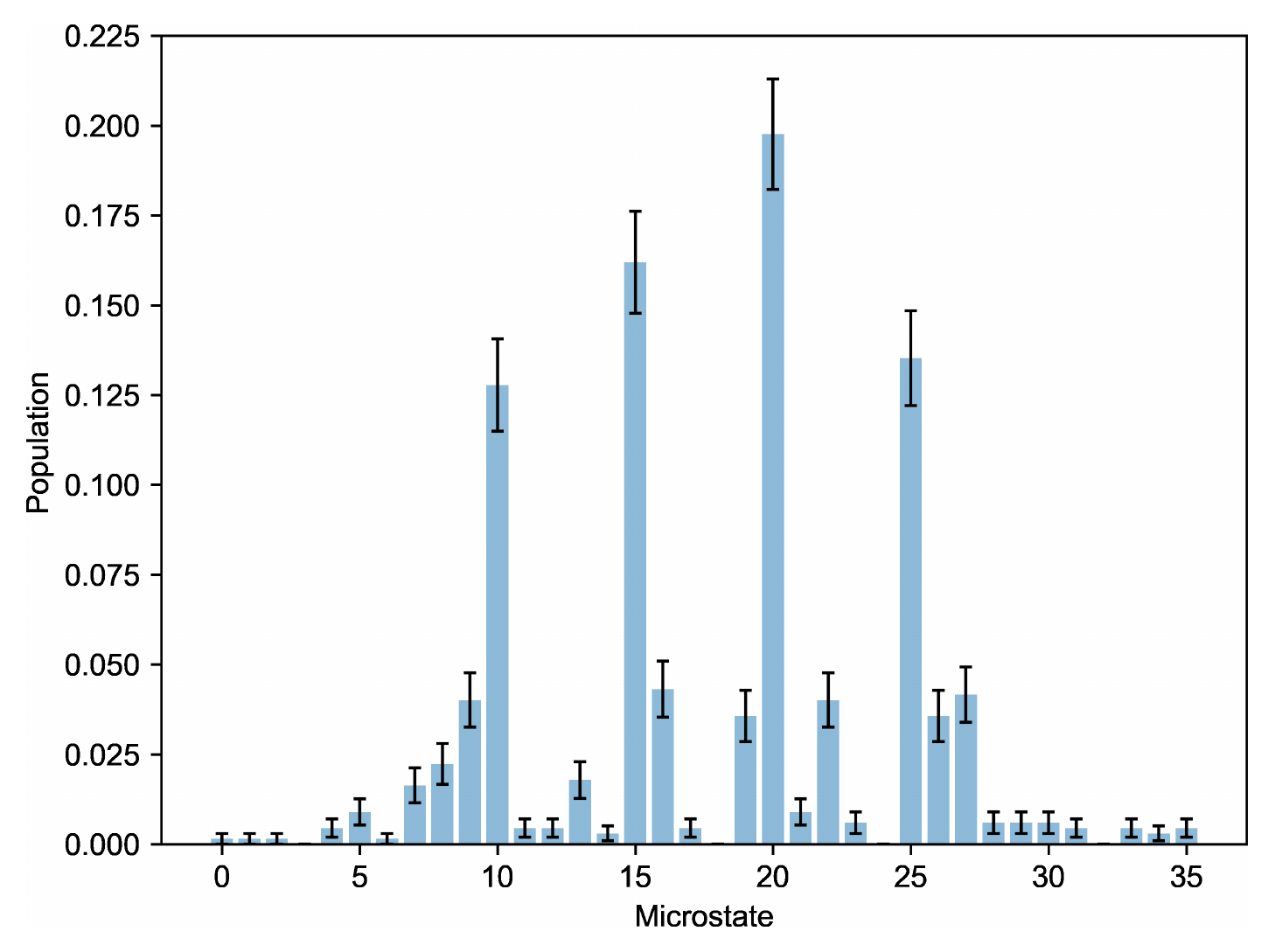}
\caption{\label{fig:fullmicro} Full microstate data for Fig.~4 of the main text without averaging to take advantage of the spin symmetry of the Hamiltonian.}
\end{figure}

In the main text of the paper, for the $2\times2$ tunneling data we average over states obtained by interchanging \upstate~and \downstate. This is because all potentials in the system are state independent. This reduces the statistical error due to the limited number of experimental shots (673 shots). Since we do not equalize the disorder entirely to zero, we do not assume any reflection symmetries that could be utilized in the ideal system to further reduce the statistical errors. We show the full microstate distribution without averaging in Fig. \ref{fig:fullmicro}, and the full labelling of microstates in Tab. \ref{tab:1}.

\begin{table}
\begin{tabular}{||c | c |  c||} 
 \hline
 Index & \upstate~population & \downstate~population  \\ [0.5ex] 
 \hline\hline
 0 & 1100 & 1100 \\ 
 \hline
 1 & 1010 & 1100 \\
 \hline
 2 & 1001 & 1100  \\
 \hline
 3 & 0110 & 1100  \\
 \hline
 4 & 0101 & 1100 \\ 
  \hline
 5 & 0011 & 1100 \\ 
 \hline
 6 & 1100 & 1010 \\
 \hline
 7 & 1010 & 1010  \\
 \hline
 8 & 1001 & 1010  \\
 \hline
 9 & 0110 & 1010  \\ 
 \hline
  10 & 0101 & 1010 \\ 
 \hline
 11 & 0011 & 1010 \\
 \hline
 12 & 1100 & 1001  \\
 \hline
 13 & 1010 & 1001  \\
 \hline
 14 & 1001 & 1001 \\ 
 \hline
  15 & 0110 & 1001 \\ 
 \hline
 16 & 0101 & 1001 \\
 \hline
 17 & 0011 & 1001  \\
 \hline
 18 & 1100 & 0110  \\
 \hline
 19 & 1010 & 0110  \\ 
 \hline
  20 & 1001 & 0110 \\ 
 \hline
 21 & 0110 & 0110 \\
 \hline
 22 & 0101 & 0110  \\
 \hline
 23 & 0011 & 0110  \\
 \hline
 24 & 1100 & 0101  \\ 
 \hline
  25 & 1010 & 0101 \\ 
 \hline
 26 & 1001 & 0101 \\
 \hline
 27 & 0110 & 0101  \\
 \hline
 28 & 0101 & 0101  \\
 \hline
 29 & 0011 & 0101 \\ 
 \hline
  30 & 1100 & 0011 \\ 
 \hline
 31 & 1010 & 0011 \\
 \hline
 32 & 1001 & 0011  \\
 \hline
 33 & 0110 & 0011  \\
 \hline
 34 & 0101 & 0011 \\ 
  \hline
 35 & 0011 & 0011 \\ [1ex] 
 \hline
\end{tabular}
 \caption{\label{tab:1} The basis microstates for the data reported in Fig.~4 of the main text. The order of the sites used corresponds to the numbering in Fig.~4(a).}

\end{table}

\section{Entropy and Temperature}

\begin{figure}[]
\includegraphics[width = 5 in]{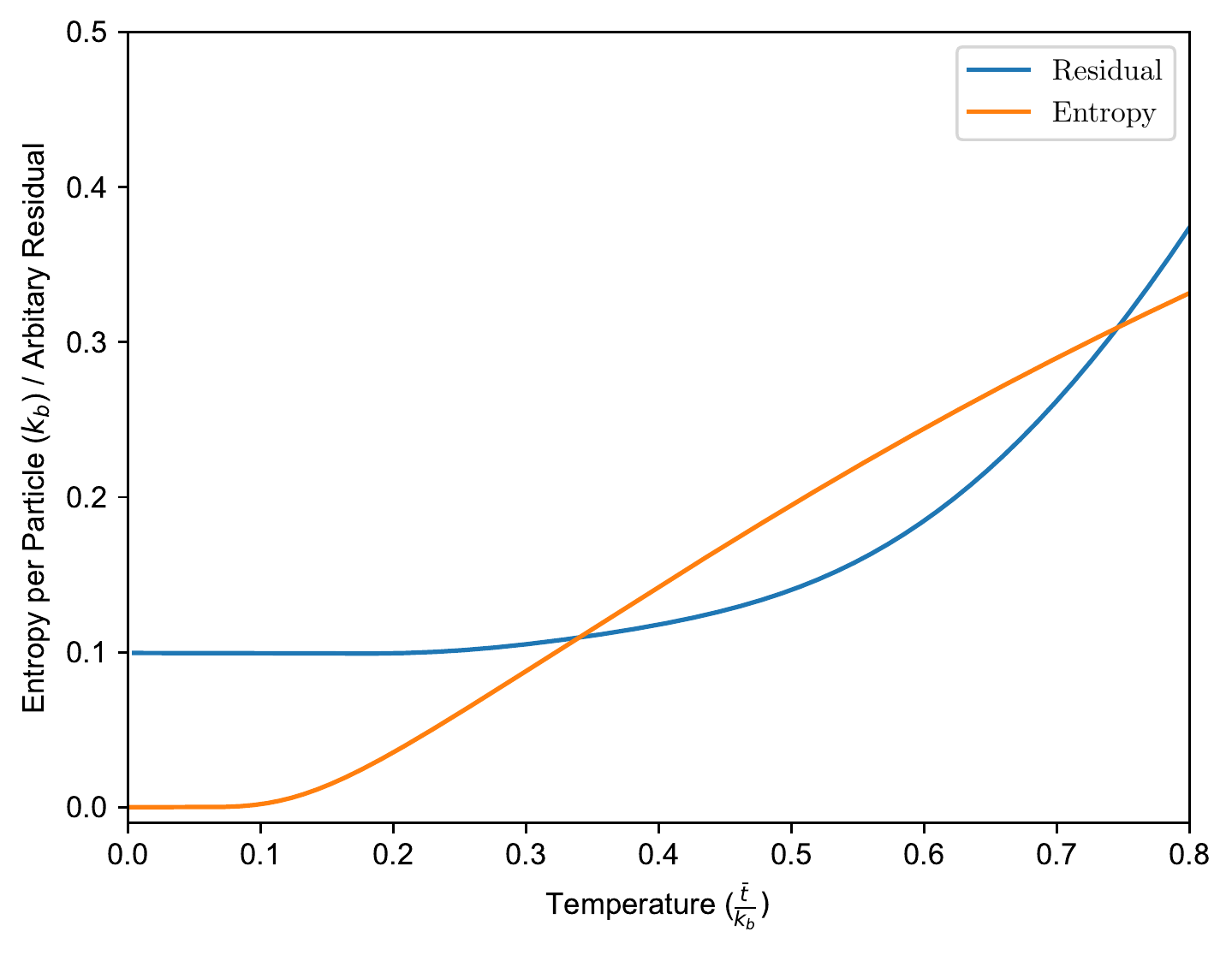}
\caption{\label{fig:entropy} Entropy and least squares residual as a function of temperature. Below around \kB$T = 0.3 \bar{t}$, the least squares residual is minimized, corresponding to an entropy range of [0,0.09] \kB~per particle.}
\end{figure}

In order to fit the temperature of our system, we use the canonical ensemble as post-selection enables us to remove number fluctuations. Thus, with free parameters temperature $T$ and $U/t$, the partition function is the sum over all energy eigenstates $\Omega$, with
\begin{equation}
    Z\left(U/t,T\right) = \sum_{\Omega} \exp \left(\frac{-E_{\Omega}(U/t)}{k_b T}\right).
\end{equation}
For each temperature and interaction, we create an expected distribution of microstates, and we compare to the experimental data using a weighted least squares fit. We find that the weighted sum of the squared residuals reaches a local minimum (which depends very weakly on temperature) at $U/t = 3.4(2)$, where the errorbar is extracted from 500 bootstrapped samples of the data. We find that the temperature fit loses sensitivity below \kB$T \sim 0.3 \bar{t}$ (Fig \ref{fig:entropy}). This corresponds to an entropy range of [0,0.09] \kB~per particle. Exact diagonalization performed by in the Python package Quspin~\cite{weinberg2019quspin} suggests that for our ramp parameters, the entropy gain would be 0.02\,\kB~per particle.

In future studies, we wish to implement larger 2D ladder systems, preparing ground states of half-filled systems by the aforementioned technique of adiabatically ramping on additional tweezer sites.
We use exact diagonalization on numerically tractable systems to estimate the many-body gaps sizes and therefore the feasibility of this approach.
For instance, on a $2\times5$ triangular lattice at half filling, at $U/t = 6$, we can perform an adiabatic ramp in 100~ms that will only lead to an increase of 0.06 $k_\mathrm{B}$ per particle.  For a $2\times6$ lattice with the same parameters, the increase is 0.07 $k_\mathrm{B}$. 
These values were obtained by calculating dynamics with experimentally realistic ramp parameters.
We are unable to perform exact diagonalization with our computing resources for larger 2D systems.

\end{document}